\documentclass[letter]{aa} 

\usepackage{epsfig}     
\usepackage{graphicx,color}     
\usepackage{amssymb}            
\usepackage{url}                
\usepackage{amsmath}            
\usepackage{rotating}                   
\usepackage{float}                      
\usepackage{textcomp}
\usepackage{epstopdf}
\usepackage{dcolumn}
\usepackage{times}
\usepackage{tabularx}
\usepackage{hyperref}
\usepackage[normalem]{ulem}
\hypersetup{
    colorlinks,
    citecolor=blue,
    filecolor=blue,
    linkcolor=blue,
    urlcolor=blue,
    menucolor=black}
\usepackage{soul} 
\usepackage[english]{babel}
\usepackage{booktabs}
\usepackage{xspace}
\usepackage{gensymb}
\usepackage{arevmath}
\usepackage{orcidlink}
\newcommand{\hrieuv}{HRI\textsubscript{EUV}\xspace}
\newcommand{\aia}[1]{#1~{\AA}\xspace}

\begin{document}
\title{Investigating coronal loop morphology and dynamics from two vantage points}

%===Authors=================
\author{
Sudip~Mandal\inst{1}\orcidlink{0000-0002-7762-5629},
Hardi~Peter\inst{1}\orcidlink{0000-0001-9921-0937},
James~A.~Klimchuk\inst{2}\orcidlink{0000-0003-2255-0305},
Sami~K.~Solanki\inst{1,3}\orcidlink{0000-0002-3418-8449},
Lakshmi~Pradeep~Chitta\inst{1}\orcidlink{0000-0002-9270-6785},
Regina~Aznar Cuadrado\inst{1}\orcidlink{0000-0003-1294-1257}, 
Udo~Sch\"{u}hle\inst{1}\orcidlink{0000-0001-6060-9078},
Luca~Teriaca\inst{1}\orcidlink{0000-0001-7298-2320},
David~Berghmans\inst{4},
Cis~Verbeeck\inst{4}\orcidlink{0000-0002-5022-4534},
F.~Auch\`{e}re\inst{5},
\and
Koen~Stegen\inst{4}
}

\institute{
Max Planck Institute for Solar System Research, Justus-von-Liebig-Weg 3, 37077, G{\"o}ttingen, Germany \\
\email{smandal.solar@gmail.com}
\and
NASA Goddard Space Flight Center, USA
\and
School of Space Research, Kyung Hee University, Yongin, Gyeonggi 446-701, Republic of Korea
\and
Solar-Terrestrial Centre of Excellence -- SIDC, Royal Observatory of Belgium, Ringlaan -3- Av. Circulaire, 1180 Brussels, Belgium
\and
Université Paris-Saclay, CNRS,  Institut d'Astrophysique Spatiale, 91405, Orsay, France
}
%===Authors end here=================

\abstract{Coronal loops serve as the fundamental building blocks of the solar corona. Therefore, comprehending their properties is essential in unraveling the dynamics of the Sun's upper atmosphere. In this study, we conduct a comparative analysis of the morphology and dynamics of a coronal loop observed from two different spacecraft: the High Resolution Imager (\hrieuv) of the Extreme Ultraviolet Imager aboard the Solar Orbiter and the Atmospheric Imaging Assembly (AIA) aboard the Solar Dynamics Observatory. These spacecraft were separated by 43{\textdegree} during this observation. The main findings of this study are: (1) The observed loop exhibits similar widths in both the \hrieuv and AIA data, suggesting that the cross-sectional shape of the loop is circular; (2) The loop maintains a uniform width along its entire length, supporting the notion that coronal loops do not exhibit expansion; (3) Notably, the loop undergoes unconventional dynamics, including thread separation and abrupt downward movement. Intriguingly, these dynamic features also appear similar in data from both spacecraft. Although based on observation of a single loop, these results raise questions about the validity of the coronal veil hypothesis and underscore the intricate and diverse nature of complexity within coronal loops.
}

   \titlerunning{Investigating coronal loop morphology and dynamics from two vantage points}
   \authorrunning{Sudip Mandal et al.}
   \maketitle
 
%=====================================
\section{Introduction} \label{sec:intro}
Coronal loops, characterized by their bright, curved, tube-like appearance, stand as some of the most easily recognizable features within the solar corona. Traditionally, these loops have been understood in terms of plasma confinement within arched magnetic field lines that extend into the low-$\beta$ corona. Depending on the wavelength at which they are observed, the plasma inside a loop is hotter and/or denser compared to the surroundings, leading to their bright appearance. Over the years, regular observations of the corona in extreme ultraviolet (EUV) and X-ray wavelengths, where these loops are most prominently visible, have led to a plethora of research on understanding their properties and evolution \citep{2014LRSP...11....4R}, including stereoscopic determination of loop geometry, density, and temperature \citep{2007ApJ...671L.205F,2008ApJ...680.1477A,2008ApJ...679..827A}. 

Among others, the shape of a coronal loop remains a topic of interest among researchers. Observations typically reveal that these loops maintain a consistent width or cross-sectional diameter along their entire length \citep{1992PASJ...44L.181K,2000SoPh..193...53K,2006ApJ...639..459L}. This is in stark contrast with our current magnetic extrapolation models that predict expansion of magnetic field with height above the solar surface. Several potential explanations have been proposed for this apparent discrepancy, including the presence of twist in the field lines \citep{2000ApJ...542..504K}, a magnetic separator that expands less \citep{2009ApJ...706..108P}, a combination of the thermal structuring of the loop and the spectral properties of the imaging instrumentation \citep{2012A&A...548A...1P} and, preferential expansion in the line-of-sight direction \citep{2013ApJ...775..120M}. However, none of these proposed solutions have been universally proven to apply to all types of loops and in different magnetic environments, such as active regions and the quiet Sun.

Another related issue concerns the cross-sectional structure of coronal loops. In EUV images, the cross-section of a loop often appears symmetric and is typically modeled using a Gaussian profile. This prompted researchers to conclude that a coronal loop possesses a circular cross-section (e.g., \citealp{2000SoPh..193...53K}). However, it is unclear why the heating would be symmetrical (a symmetrically spreading avalanche of nanoflares is one possibility; \citealp{2023ApJ...942...10K}), and therefore, why a loop would have a circular cross-section. Nonetheless, it is important to note that, like many other studies of the solar corona, assessments of loop properties are also affected by the optically thin nature of the coronal emission. Features in the background or foreground contaminate measurements \citep{2021ApJ...913...56M}, although results about constant loop width may still hold true \citep{2008ApJ...673..586L}.

%++++++++++++++++++++++++++++++++++++++++
\begin{figure*}[!ht]
\centering
\includegraphics[width=0.90\textwidth,clip,trim=0cm 0cm 0cm 0cm]{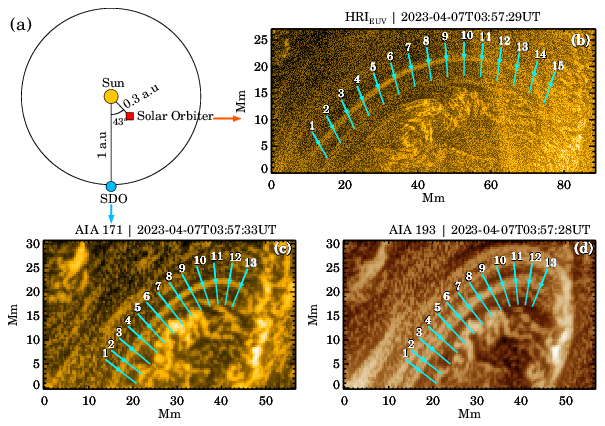}
 \caption{An overview of the event. Panel-a depicts the relative position of the two spacecraft, SDO and Solar Orbiter, whose data are used in this study. Panel-b shows the loop under study in the \hrieuv image, while panel-c and d show the same loop, but as seen in AIA \aia{171} and \aia{193} channels, respectively. The cyan lines highlight the locations of the artificial slits that are used to generate space-time maps (shown in Figs.~\ref{fig:xt},~\ref{fig:width}, ~\ref{fig:oscillation} and ~\ref{fig:xt_all}). The arrow in the center of each slit indicates the direction of increasing distance along the slit. Images in panel-b, c and d are unsharp-masked for improved visibility of the loop.
Movie \url{https://drive.google.com/file/d/1Baoaaq6gPXD_LZdAOjjScEXCkonj18gl/view?usp=sharing}.}
\label{fig:context}
\end{figure*}
%++++++++++++++++++++++++++++++++++++++++

An alternative interpretation of the cross-sectional loop profile is the `coronal veil'  hypothesis \citep{2022ApJ...927....1M}. According to this, loops are a line of sight effect of warped sheets of bright emission. This scenario is similar to how wrinkles appear in a veil. However, similar to other aspects of this model picture, to evaluate the cross-sectional shape of loops, it is imperative to observe the same loop from different vantage points, resulting in two distinct line-of-sight integrations.

 In this study we compare the dynamics and morphology of a coronal loop viewed from two spacecraft that, at the time of the observations analysed here, subtended a 43{\textdegree} angle at the Sun. We use co-temporal high-resolution EUV images of the corona taken from the Solar Dynamics Observatory \cite[SDO;][]{2012SoPh..275....3P} and the Solar Orbiter spacecraft \citep{2020A&A...642A...1M}. This approach enables us to further investigate the properties of the loop in connection with the `coronal veil' hypothesis.

%`````````````SECTION:DATA`````````````````
\section{Data} \label{sec:data}
%++++++++++++++++++++++++++++++++++++++++++
\begin{figure*}[!ht]
\centering
\includegraphics[width=0.80\textwidth,clip,trim=0cm 0cm 0cm 0cm]{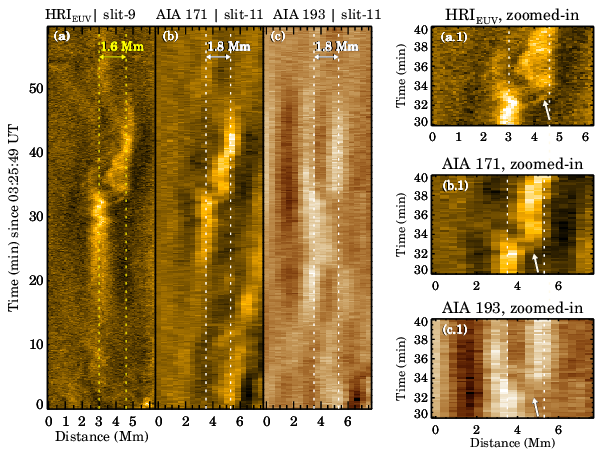}
\caption{Representative examples of space-time (x-t) maps derived from the \hrieuv (panel-a), AIA~\aia{171} (panel-b) and AIA~\aia{193} (panel-c) image sequences. Zoomed-in versions of these maps, between t=30 min and 40 min, are presented in panels-a.1, b.1, and c.1, respectively. Arrows in these zoomed-in panel point to the slanted ridge created by the thin strand. The dotted vertical lines in these panels outline the shift of the loop as judged visually.}
\label{fig:xt}
\end{figure*}
%++++++++++++++++++++++++++++++++++++++++++
%++++++++++++++++++++++++++++++++++++++++++
\begin{figure*}[!ht]
\centering
\includegraphics[width=0.85\textwidth,clip,trim=0cm 0cm 0cm 0cm]{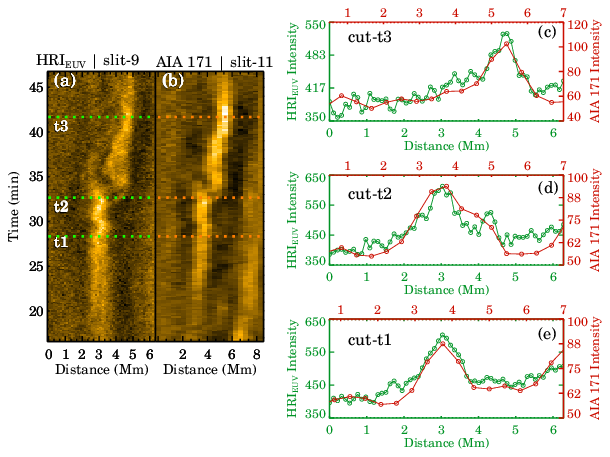}
\caption{Comparing loop widths from \hrieuv and AIA. Panels-a and b show x-t maps from \hrieuv and AIA, respectively. Intensity (DN s$^{-1}$) along the respective coloured dashed lines (marked with `t') are plotted in panels c-e. Panel-c shows the derived curves from \hrieuv (green curve) and AIA~\aia{171} (red) data, at t=$\mathrm{t3}$. The curves from t=$\mathrm{t2}$ and $\mathrm{t1}$ are shown in panels-d and e, respectively. 
}
\label{fig:width}
\end{figure*}
%++++++++++++++++++++++++++++++++++++++++++

We used extreme ultraviolet (EUV) images taken on 2023-04-07 by Solar Orbiter and SDO. We utilized EUV images from the High Resolution Imager (\hrieuv; taken via the 174~{\AA} bandpass) of the Extreme Ultraviolet Imager \cite[EUI;][]{2020A&A...642A...8R} which samples plasma with a temperature of around
T$\approx$1 MK. This \hrieuv dataset\footnote{Part of the SolO/EUI Data Release 6.0 \citep{euidatarelease6} and available publicly.} has a cadence of 10\,seconds, lasted for one hour and has an image scale of 0.492\arcsec\,pixel$^{-1}$. 
On that day the Solar Orbiter was about 0.3 astronomical units (au) away from the Sun, meaning that the \hrieuv\ images have a plate scale of 108\,km\,pixel$^{-1}$ on the Sun. Solar Orbiter was about 43{\textdegree} away from the Sun-Earth line. Additionally, we combined the \hrieuv data with full disc EUV images from the Atmospheric Imaging Assembly~\cite[AIA;][]{2012SoPh..275...17L} aboard the Earth orbiting SDO. Specifically, we analyzed data from the 171~{\AA} (sensitive to plasma of 0.8~MK), 193~{\AA} (1.6~MK), and 211~{\AA} (2.0~MK) AIA passbands, each with a cadence of 12\,seconds and a plate scale of 0.6\arcsec\,pixel$^{-1}$ (corresponding to 435\,km\,pixel$^{-1}$ on the Sun). While the \hrieuv data have almost four times better spatial resolution than the AIA data, both datasets have similar temporal resolution. Lastly, while comparing the \hrieuv and AIA images, we took into account the difference in light propagation time from the Sun to Solar Orbiter, which was 0.3\,au away from the Sun, and to the SDO, which was 1\,au away. All the time-stamps quoted here are the times as measured at Earth.

\section{Results}

We are focusing on a coronal loop situated on the northwest side of the active region $\texttt{NOAA AR13270}$, which is at the center of the \hrieuv field-of-view (see Fig~\ref{fig:full_fov} in the appendix). It is important to note that Solar Orbiter was at an angle of 43{\textdegree} with SDO when the observation was made, as shown in Fig.~\ref{fig:context}a. By combining images from AIA and \hrieuv we were able to obtain a stereoscopic view of the loop and its dynamics.

\subsection{Comparing Loop dynamics}

In the animation shown in Fig~\ref{fig:context},  we find a loop (or a group of threads) appearing first around 03:45 UT. Over time, it gradually becomes brighter and undergoes various dynamic changes. This evolution appears similar in both \hrieuv and AIA data, even though the latter instrument is at an angular distance of 43\textdegree\ from the former. To  quantitatively analyze the loop's evolution, we placed multiple artificial slits along its length, as shown in Fig.~\ref{fig:context}b, c, and d. Through these slits, we aim to capture the loop's dynamics, including any oscillations that occur perpendicular to the loop. It is important to clarify that, since our goal is to study the overall characteristics of the loop, we do not precisely align these artificial slits in exactly the same positions in both images. Instead, we aim to place them nearby, as establishing a pixel-level correspondence between these two datasets is challenging.

The space-time (x-t) maps for a set of artificial slits are presented in Fig.~\ref{fig:xt}. We will first focus on the \hrieuv x-t map (panel-a). The loop (at x=3 Mm) gradually brightens up starting from t=15 min. Then, starting at t=27 min, it undergoes transverse oscillations as indicated by the sinusoidal pattern in the map (also visible in the animation). While the oscillations were present, a thin thread-like structure appears to separate from the loop and to move away. Panel-a.1 presents a closer view of this segment, with the thin thread (that gets separated) highlighted by a white arrow. The thread stops moving after traveling almost 1.6 Mm along the slit in just one minute. Interestingly, the entire loop bundle (from which the thin thread is detached) is also observed to be displaced (ending at around the x=4.6 Mm mark) by almost the same distance of 1.6 Mm as the thin thread. The extent of this motion is highlighted by two vertical dashed lines in panel-a. Again, this movement also happens over a time scale of one minute. Eventually, the shifted loop fades away gradually over time. In summary, the \hrieuv images display a loop rapidly moving in the transverse direction by $\sim$1.6 ~Mm within one minute.

Let us turn our attention to the AIA images. In Figure~\ref{fig:xt}, panels b and c show the x-t maps for the \aia{171} and \aia{193} channels, respectively\footnote{We compared slit-11 of AIA with slit-9 of \hrieuv (see Fig.\,\ref{fig:context}) after visually verifying their proximity in location and the resemblance in the evolution of loops in their respective x-t maps. Furthermore, evolution of the loop appears similar in other nearby AIA slits as well (see Fig.~\ref{fig:xt_all}).}. The loop evolution in the \aia{171} channel appears similar to that in the \hrieuv, although the lower spatial resolution of AIA is noticeable in the map. Nevertheless, we can also identify the thin thread in this map (panel b.1), primarily due to our prior knowledge about it from \hrieuv data. Remarkably, we found the displacement of the loop (located at x=3.5 Mm) in the \aia{171} channel to be similar (1.8 Mm, as highlighted by two vertical dashed lines) to that of \hrieuv (1.6~Mm), despite the two instruments being 43{\textdegree} apart. This result suggests that the plane of motion of the loop (and the thin thread) is roughly perpendicular to the solar surface.

Interestingly, the \aia{193} channel map (panel c of Fig.~\ref{fig:xt}) not only exhibits similarities, but also significant differences when compared to the \aia{171} and \hrieuv maps. For example, between t=15 min and t=22 min, the loop (located at x=3.5 Mm) appears significantly brighter in the \aia{193} map than in the \aia{171} map. Furthermore, at x=5.3 Mm (the second vertical line), a bright loop is visible in the \aia{193} map, while we find no such structure in the \aia{171} map. In contrast, between t=22 min and 32 min, the loop is clearly discernible in the \aia{171} map, but appears somewhat blurry in the \aia{193} map. What makes this even more intriguing is the comparison of the times after which the loop suddenly moves downwards in the \aia{171} data. In the \aia{193} map, we can see a loop exactly where it eventually settles in the 171 map after the movement (i.e., at x=5.3). However, we also continue to observe a loop in the \aia{193} map at the position\footnote{By comparing the location of the second dashed vertical line, it appears that there is a one-pixel shift in the position of the \aia{193} loop relative to the \aia{171} loop.} 
where the loop was previously before the movement (x=3.5). Therefore, we have a scenario where the loop moves downwards in the \aia{171} map, while in the \aia{193} map, we see two loops - one at the shifted loop position and another where the loop was before the shift.

\subsection{Comparing loop morphology}
Here we analyze the shape and appearance of the loop as seen through \hrieuv and AIA~\aia{171} images. These analyses were performed on the raw data, not on the edge-enhanced images.

At first glance, the shape and evolution appear quite similar in \hrieuv and AIA (Fig.~\ref{fig:xt}). To quantitatively compare the two, we examined the width (via cross-sectional intensity) of the loop at three different times: before, during, and after the sudden movement of the loop, as highlighted in Fig.~\ref{fig:width}. The loop width appears similar in \hrieuv and \aia{171} at all three instances. This also provides information about the shape of the loop cross-section, a topic that is still actively debated in the community \citep{2000SoPh..193...53K,2000ApJ...542..504K,2013ApJ...775..120M,2021ApJ...919...47W,2023arXiv231007102U}. If the loop has an elliptical cross-section, we expect to see changes in the measured cross-section values (and therefore in cross-sectional intensities) when viewed from 43\textdegree~apart. However, as revealed in Fig.~\ref{fig:width}, we do not find any such difference. Therefore, we conclude that the loop's cross-section is nearly circular, consistent with previous studies such as by \citet{2000SoPh..193...53K} and \citet{2020ApJ...900..167K}. In addition to this, Fig.~\ref{fig:width}d shows that the two structures, the parent loop and the thin thread, are well resolved in \hrieuv (the green curve). Interestingly, despite having four times coarser spatial resolution than \hrieuv, AIA (the red curve) also captured the small thread, albeit only just. Moreover, it is also evident that without the assistance from the \hrieuv image, one would likely not consider the \aia{171} feature as a signature of the thread.

\begin{figure}[!h]
\centering
\includegraphics[width=0.40\textwidth,clip,trim=0.1cm 0cm 0cm 0cm]{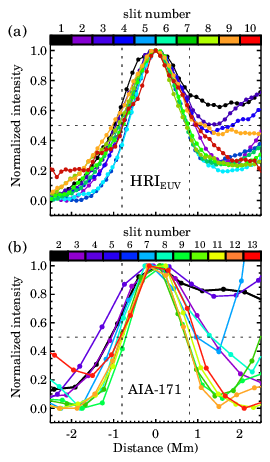}
\caption{Variation of loop width along its length. Panel-a shows the normalized \hrieuv intensities calculated along different slits (highlighted via the colorbar). The same but for the AIA~\aia{171} data is shown in panel-b. The time at which these loop intensities were derived is identical to that shown in Fig.~\ref{fig:context}. Each curve is adjusted to ensure that its peak lies at x=0 Mm. The dotted vertical and horizontal lines in each panel act as references to approximate the Full Width at Half Maximum (FWHM). AIA curves from slit-5 and slit-6 are not displayed due to significant contamination from the nearby moss-type structure.
}
\label{fig:width_along_length}
\end{figure}

Next, we examined the loop width along the length of the loop. In Fig.~\ref{fig:width_along_length} we show the loop intensities along different slits which are placed perpendicular to the loop's length as shown in Fig.~\ref{fig:context}. The time at which these loop intensities were derived is identical to that shown in Fig.~\ref{fig:context}. To get a better understanding of the overall behavior, we applied running averages on the curves. We used 8 pixels in \hrieuv and 2 pixels in AIA, taking into account the four-fold resolution difference between these two instruments. This smoothing process effectively mitigated minor fluctuations. Moreover, in order to avoid low signal-to-noise, we limit our analysis of the \hrieuv data to the region spanning from slit 1 to slit 10. Regardless, through these slits, we cover more than half of the loop's total length. Our analysis yielded two crucial outcomes: a) The loop's width remains remarkably consistent along its length, from both spacecrafts. This characteristic is more prominently captured in \hrieuv data, owing to its superior spatial resolution. Furthermore, the proximity of a moss-type structure adjacent to the AIA loop section near slits-4 to 7 affects the shape of the intensity curves from these slits, resulting in distortion and broadening. b) The loop's width, as observed from both \hrieuv and AIA perspectives, aligns along its length (the full width at half maxima (FWHM) is roughly 1.6 Mm in both datasets). Because the loop is not completely aligned with the plane spanned by the vantage points of the spacecraft\footnote{It is evident from Fig.~\ref{fig:context} that the loop runs diagonally from South-East to North-East with a considerable curvature. Therefore, the two lines of sight are at an angle with the loop's plane.}, these results suggest that the loop maintains a nearly circular cross-section throughout its entire extent.

In summary, based on our analysis of a coronal loop viewed from two spacecraft at a 43{\textdegree} angle separation, we conclude that the loop's width and its structural evolution exhibit remarkable similarities. This finding challenges the viability of a coronal veil-like scenario as an explanation, at least in the context of this specific case.

%-------------------
%------------------

%-------------------
\begin{figure*}[!ht]
\centering
\includegraphics[width=0.99\textwidth,clip,trim=0cm 1cm 0cm 0cm]{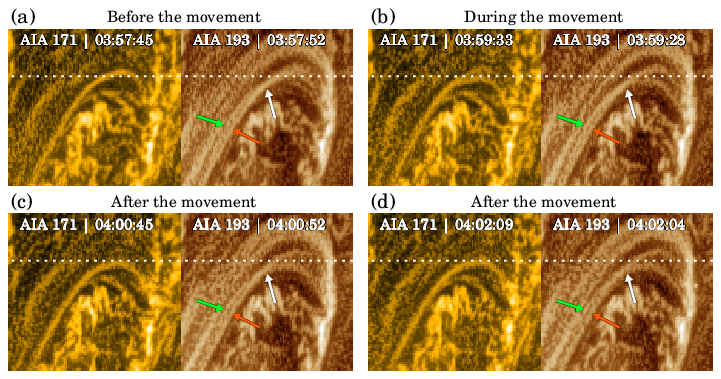}
\caption{ Snapshots from the AIA image sequence. Panel-a presents snapshots from the \aia{171} channel (left) as well as from the \aia{193} channel (right), before the loop started moving downward in the \aia{171} data. The same but for instances during and after the movement are shown in panel-b and panels-c-d, respectively. The dotted white line in every panel serves as a fiducial marker to spot the loop displacement in the \aia{171} images. In each panel, on top of the \aia{193} image, the green and red arrows point to the two separate threads, while the white arrow points to the location where they appear to cross each other. See Sect.~\ref{sec:projection} for details.}
\label{fig:two_loop}
\end{figure*}
%------------------

\section{What explains the observed loop dynamics?}

Our analysis has brought forth a series of intriguing questions regarding the dynamics of the loop. These include: (i) The origin of the downward motion observed in both the \hrieuv and AIA \aia{171} images; (ii) The rationale behind the consistent shifts in the loop as observed by two spacecraft positioned 43 degrees apart; (iii) The factor(s) responsible for the loop's simultaneous appearance in both AIA channels at times, while at other instances, it appears in one (\aia{171}) and remains absent in the other (\aia{193}); (iv) The potential role of the thin strand in shaping the overall evolution of the system. In the following sections we explore possible explanations to these.

\subsection{Projection effects?}\label{sec:projection}

Upon careful examination of the images from the AIA~\aia{193} and AIA~\aia{171} channels, it appears that the presence of two loops in the former as compared to one loop in the latter may be attributed to projection effects. In fact, after reviewing the animation associated with Fig.~\ref{fig:context}, it becomes apparent that there were, in fact, two loops present from the beginning. Nevertheless, these two loops were oriented in such a manner that, along most of their length, they appeared as a unified and cohesive structure. It is only at the apex where these two structures diverge, becoming discernible as distinct loops. To further support this conclusion, we have included four snapshots in Fig.~\ref{fig:two_loop}, where we have highlighted the two loops with red and green arrows and the possible location of crossing with white arrows. While this crossing structure may appear to suggest loop braiding, it is much more likely to be a mere projection effect. This is because apparently braided and interacting strands within a loop bundle appear to exhibit rapid intensity variations \citep{2022A&A...667A.166C}, which is not observed in our case. Additionally, upon reviewing the AIA x-t maps from slits-6, 8 and 9 as shown in Appendix B, it becomes evident that the two loops in the AIA~\aia{193} channel were indeed present from the beginning. However, these results do not provide an explanation for the sudden downward movement of one loop or the abrupt disappearance of the other loop from the \aia{171} channel while remaining visible in the \aia{193} channel.

\subsection{Heating or cooling?}\label{sec:heat_cool}

%-------------------
\begin{figure*}[!ht]
\centering
\includegraphics[width=0.85\textwidth,clip,trim=0cm 0.6cm 0cm 0.3cm]{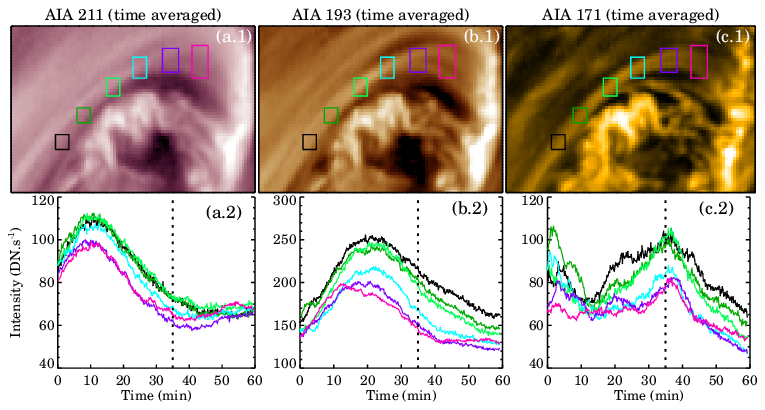}
\caption{Evolution of the loop intensities in different AIA channels. Panel-a.1 shows the time-averaged \aia{211} image with the boxes of different colors highlighting the locations from where the average intensities (DN.s$^{-1}$) shown in Panel-a.2 are derived. The same but for the \aia{193} and \aia{171} channels are shown in panels (b.1, b.2) and (c.1, c.2), respectively. The vertical line in each panel of the bottom row indicates the timestamp when the loop is first seen to move downward in the \aia{171} channel.}
\label{fig:lc}
\end{figure*}
%------------------

The visibility of a feature in a given AIA passband depends on its temperature and/or density. If we assume that the loop density remains approximately constant during the observation, the intensity fluctuations can then be attributed solely to the change in loop temperature. Therefore, if the loop is visible in the \aia{193} channel but not in the \aia{171} channel, it could be because the loop is too hot to be captured in that particular AIA passband. Conversely, if the loop is present in two passbands at the same time, it may indicate that the loop is either multi-thermal or its temperature falls within the response function of both passbands. To understand the evolution of the loop, we examine the intensities at different positions along its length using boxes that cover its lateral extension, as shown in Fig~\ref{fig:lc}. Light curves from the \aia{211} (panel-a.2 of Fig.~\ref{fig:lc}), \aia{193} (panel-b.2) and \aia{171} (panel-c.2), peak progressively at later times, implying that the loop is undergoing cooling (see Appendix~\ref{appendix:c} for more on cooling time). Interestingly, the shape of the \aia{171} curves (panel-c.2) are rather steep (near their maximums) compared to the other two channels. Moreover, the vertical dotted line that marks the time when we first observed the downward loop movement in the \aia{171} images, coincides with the peak of the light curve in panel-c.2. This means that the loop starts to cool in \aia{171} channel (rather steeply) at the same time as it starts moving downward. At this point, we cannot determine if this is anything more than a coincidence.

This overarching cooling scenario introduces further complexities to an already complicated evolutionary sequence. Previously, the presence of the loop in \aia{193} images and its absence in the \aia{171} images (panels-b and c of Fig.~\ref{fig:xt}) might have been attributed to a heating event, such as via reconnection. However, this explanation appears less probable now, given the ongoing cooling of the system. Nevertheless, it remains plausible that a localized, small-scale heating event did occur at that specific location, but it went undetected in the AIA (and \hrieuv) images.

\subsection{Oscillation induced reconnection?}

\begin{figure}[!ht]
\centering
\includegraphics[width=0.40\textwidth,clip,trim=0cm 0cm 0cm 0cm]{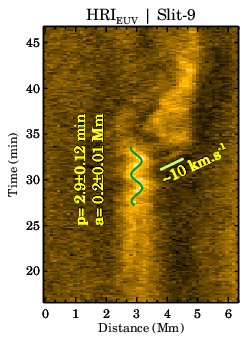}
\caption{ Oscillations in the \hrieuv x-t map. The green curve outlines the fit to the observed transverse oscillations. Derived parameters are printed on the panel. The green line shows the best fit to the slanted ridge above it. The speed, measured through the slope of the dashed line, is also printed on the panel.}
\label{fig:oscillation}
\end{figure}

Prior to the detachment of the thin thread, the \hrieuv x-t map (Fig.~\ref{fig:xt}a) displays signatures of transverse oscillations. These oscillations do not exhibit any noticeable change in their amplitude over the two cycles we observe. We fit the observed oscillation as shown in Fig~\ref{fig:oscillation} and calculated the oscillation period ($\mathrm{p}$) to be 2.9 min, with the amplitude ($\mathrm{a}$) being 0.2 Mm. These parameters are similar to typical decayless kink oscillations that are found in active region loops as reported in \citet{2015A&A...583A.136A} and \citet{2022A&A...666L...2M}. Curiously, even after the small thread was detached, the parent loop continued to oscillate. This aspect suggests that the transverse oscillation and thread detachment are separate and unrelated events. Consequently, whether the oscillations played a role in triggering the downward movement of the loop or not, remains a speculation at this point.

%-------------------
%------------------

It is possible that the observed transverse oscillations have induced reconnection owing to the small-angle misalignment between threads of the loop. As a result, some field lines of the parent loop were pushed sideways, resulting in the appearance of the thread. However, the speed at which the thread moves away (10~km.s$^{-1}$, see Fig.~\ref{fig:oscillation}) is significantly lower than the typical Alfv\'en speed ($\sim$1000~km.s$^{-1}$). It is however possible that the small angle misalignment of the field leads to a smaller field component and subsequently smaller Alfv\'en speed. Therefore, the thin thread is indeed a product of magnetic reconnection as the heat deposited in such a case would quickly get distributed along the guide field. Further investigations are needed to confirm this hypothesis.

\section{Summary and conclusion}
Using high-resolution images from \hrieuv\ aboard Solar Orbiter and AIA aboard SDO, we analyzed the evolution of a coronal loop from two vantage points that are 43{\textdegree} apart. Below we summarize our main findings:\\

    % \item
    {\it Uniform cross-sectional shape and consistency across vantage points:} When measured through both \hrieuv and AIA~\aia{171} images, the width of the loop appears to be similar. This similarity remains consistent throughout the evolution of the loop and along its entire length. These findings suggest that the loop is essentially circular in cross-section. Additionally, it does not support the coronal veil hypothesis, which predicts that the loop's morphology would appear different when viewed from two different perspectives. However, we are aware of the limitations of our dataset, specifically that the alignment of the two lines of sight (referring to directions, not angular difference) is sub-optimal. Ideally, the best-case scenario would involve the two lines-of-sight lying in a plane perpendicular to the loop's plane. However, in the current dataset, both lines-of-sight roughly align within the loop's plane (for the most part), possibly resulting in a smaller capability to distinguish between the two dimensions of the cross-section.\\

    % \item 
    {\it Atypical loop evolution:} As seen through \hrieuv, the loop undergoes a unique evolutionary sequence, initially displaying transverse oscillations before a slender thread-like structure detaches from the primary loop. Following this, the main loop also shifts, traversing a distance of around 1.6 Mm within a matter of minutes.\\

    % \item
    {\it Unknown driving mechanism(s):} Currently, the reason(s) behind the observed loop evolution remains unclear. Possible scenarios, including projection effects, heating or cooling events, and wave-induced reconnection, do not appear to be the cause in this particular event. Therefore, we require additional information, either from another similar observation or through numerical models, to gain a better understanding of the evolution.\\

In conclusion, our study highlights the importance of multi-perspective observations in unraveling the complex behaviors of coronal loops. While our findings of unexpected consistency in loop characteristics across divergent viewing angles challenge the validity of the coronal veil theory, we cannot make a conclusive statement regarding its applicability (or lack thereof) to all coronal loops. In fact, \citet{2022ApJ...927....1M} also found a mix of veil-like and thin flux tube-like structures in their work, highlighting the complexity of the problem. A statistical study that includes a variety of loops will be helpful in this regard.

%=====================================
\begin{acknowledgements}
Solar Orbiter is a space mission of international collaboration between ESA and NASA, operated by ESA. The EUI instrument was built by CSL, IAS, MPS, MSSL/UCL, PMOD/WRC, ROB, LCF/IO with funding from the Belgian Federal Science Policy Office (BELSPO/PRODEX PEA 4000112292 and 4000134088); the Centre National d’Etudes Spatiales (CNES); the UK Space Agency (UKSA); the Bundesministerium für Wirtschaft und Energie (BMWi) through the Deutsches Zentrum für Luft- und Raumfahrt (DLR); and the Swiss Space Office (SSO). We are grateful to the ESA SOC and MOC teams for their support. Solar Dynamics Observatory (SDO) is the first mission to be launched for NASA's Living With a Star (LWS) Program. The data from the SDO/AIA consortium are provided by the Joint Science Operations Center (JSOC) Science Data Processing at Stanford University. L.P.C. gratefully acknowledges funding by the European Union (ERC, ORIGIN, 101039844). Views and opinions expressed are however those of the author(s) only and do not necessarily reflect those of the European Union or the European Research Council. Neither the European Union nor the granting authority can be held responsible for them. The work of JAK was supported by the GSFC Heliophysics Internal Scientist Funding Model competitive work package program.
\end{acknowledgements}

%======================================
\bibliography{loop_split_ref}
\bibliographystyle{aa}

%++++++++++++++++++++++++++++++++++++++++
\begin{appendix}

\section{Context image}
\begin{figure*}[!ht]
\centering
\includegraphics[width=0.85\textwidth,clip,trim=0cm 0.5cm 0cm 0.6cm]{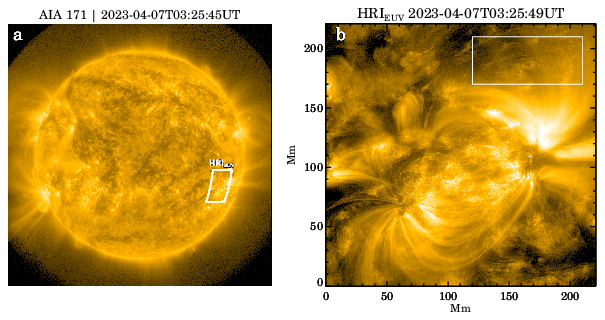}
\caption{Full FOVs of AIA (panel-a) and \hrieuv (panel-b) datsests. The white rectangle in panel-a represents the \hrieuv FOV, while the rectangle in panel-b outlines the region where the studied loop appears.}
\label{fig:full_fov}
\end{figure*}

In Figure~\ref{fig:full_fov}, we provide an overview of the entire field-of-view (FOV) captured by the AIA (shown in panel-a) and \hrieuv data (panel-b). In Figure~\ref{fig:full_fov}b, it becomes apparent that the loop we are focusing on (indicated by the white rectangle) is situated at a distance from the active region and in close proximity to a dark, filament-like structure.
%-------------------
%------------------

\section{AIA x-t maps}
\begin{figure*}[!ht]
\centering
\includegraphics[width=0.95\textwidth,clip,trim=0cm 0cm 0cm 0cm]{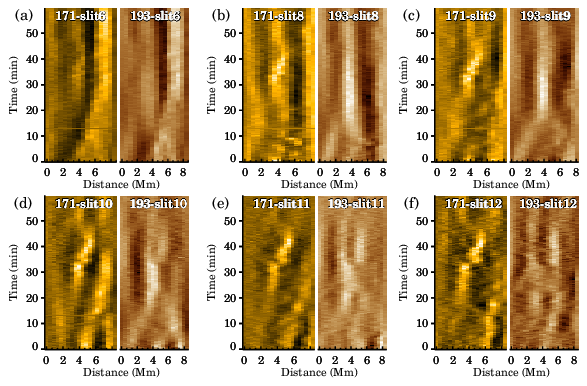}
\caption{Further examples of AIA x-t maps. Each panel contains two maps, with the left one showing data from \aia{171} and the right one showing data from \aia{193}. The slits used to create these maps are displayed on top of each panel.}
\label{fig:xt_all}
\end{figure*}

As discussed in Section~\ref{sec:projection}, the \aia{193} images reveal the existence of two loops that are positioned in such a way as to create the illusion of a single structure along the majority of their length. In Figure~\ref{fig:xt_all}, we display x-t maps obtained from various slits, illustrating that as we progress from the loop's footpoint towards its apex, the two loops gradually become more distinct and discernible.
%-------------------
%------------------

\section{Estimation of cooling times}\label{appendix:c}
\begin{figure*}[!ht]
\centering
\includegraphics[width=0.95\textwidth,clip,trim=0cm 3.2cm 0cm 0cm]{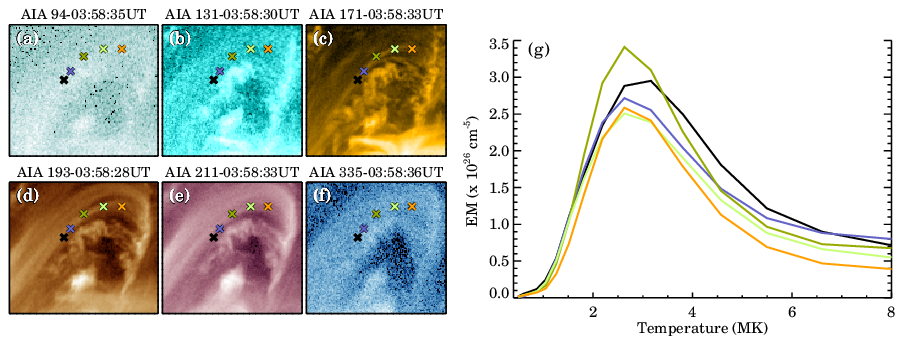}
\caption{Emission measure analysis of the loop. Panels a to f show the loop (outlined by the colored cross symbols) in six EUV passbands of AIA. The emission measure curves derived at the locations of those crosses are shown in panel g. }
\label{fig:dem}
\end{figure*}

In Section~\ref{sec:heat_cool}, we found that the loop cools down gradually over time and the observed time to cool from \aia{211} (emission peaks at t=10 min) to \aia{171} (emission peaks at t=35 min) is 25 min. We calculate here the theoretical value of the cooling time by estimating the radiative and conductive losses. We start that by first estimating the electron density as \(n= \sqrt{EM/fd}\), where EM=emission measure, f=filling factor and, d=diameter of the loop. We calculated the EM\footnote{It is obtained at time t2, as indicated in Fig.~\ref{fig:width}. Nevertheless, the EM values obtained at other times e.g., t1 and t3, exhibit considerable similarity.} by following the inversion method of \citet{2015ApJ...807..143C} and the results are presented in Fig.~\ref{fig:dem}. The average EM value as estimated along the length of the loop, is approximately 3$\times$10$^{26}$ cm$^{-5}$ at the peak temperature of 2.5 MK (see Fig.~\ref{fig:dem}g). We set the filling factor (f) to be 1 while, the diameter of the loop (d) is set as 1.6 Mm (see Fig.~\ref{fig:width}). Using these values, we estimate the loop density (n) as 1.36$\times$10$^9$ cm$^{-3}$.

The radiative cooling time ($\tau_r$) is calculated as:
\begin{equation}\label{eqn1}
    \tau_{r}=\frac{\frac{3}{2}P}{n^2\Lambda_{0}T^b}
\end{equation}

where P and T represent pressure and temperature, while $\Lambda_{0}$ is the optically thin radiative loss factor. Using the ideal gas law, P=2nkT, where k is the Boltzman's constant, into Eqn.~\ref{eqn1} we obtain

\begin{equation}\label{eqn2}
    \tau_{r}=\frac{3k}{\Lambda_{0}}
    ~ n^{-1} T^{1-b}
\end{equation}

Values of $\Lambda_{0}$ and b are set to 3.53$\times$10$^{-13}$ and -$\frac{3}{2}$, following Eqn.3 of \citet{2008ApJ...682.1351K}.

Next, the conductive cooling time ($\tau_c$) is calculated as: 
\begin{equation}\label{eqn3}
\begin{split}
    \tau_{c} & =\frac{\frac{3}{2}P}{\frac{2}{7}\kappa_{0}\frac{T^{7/2}}{L^2}}\\
    & =\frac{21}{2}\frac{\kappa}{\kappa_{0}}L^{2}nT^{-5/2}
\end{split}
\end{equation}

where L is the temperature scale length, typically taken to be the loop half length and $\kappa_{0}$=10$^{-6}$. In our case, L$\approx$50 Mm. 

Inserting the values of n, L, d and T from the observation we arrive at $\tau_c=$5.1$\times$10$^3$s and $\tau_r$=8.5$\times$10$^3$s. The total cooling time ($\tau$) from conduction and radiation is expressed as \(\tau=(\tau_c^{-1}+\tau_r^{-1})^{-1}\). Therefore, we obtain $\tau$=3.1$\times$10$^3$s =53~min (the observed cooling time is 25 min). Taking into consideration the numerous approximations we made to arrive to this value, we conclude that the observed and theoretical values are essentially consistent. The fact that the conductive and radiative cooling times are similar suggests that the loop is at the stage of cooling where evaporation is transitioning to draining. This is the time of maximum density and minimum density variation, lending support to our assumption of constant density in producing the light curves of Fig.~\ref{fig:lc}.

%-------------------
%------------------
\end{appendix}

\end{document}